\documentclass[%
 reprint,
 superscriptaddress,
 amsmath,amssymb,
 aps,
 pra,
]{revtex4-2}

\usepackage{graphicx}
\usepackage{dcolumn}
\usepackage{bm}
\usepackage{braket}
\usepackage{framed}
\usepackage{setspace}
\usepackage[colorlinks=true,linkcolor=blue, urlcolor=blue, citecolor=blue]{hyperref}
\usepackage{ulem}
\usepackage{soul}

\begin{document}

\preprint{APS/123-QED}

\title{Telecom-band Hyperentangled Photon Pairs from a Fiber-based Source}

\author{Changjia Chen}
\email{changjia.chen@mail.utoronto.ca}
\author{Calvin Xu}
\author{Arash Riazi}
\author{Eric Y. Zhu}
\author{Alexander Greenwood}
\affiliation{Dept. of Electrical and Computer Engineering, University of Toronto, 10 King's College Rd., Toronto, M5S 3G4, Canada}
\author{Alexey V.Gladyshev}
\affiliation{Fiber Optics Research Center, Russian Academy of Sciences, 38 Vavilov Street, 119333 Moscow, Russia}
\author{Peter G. Kazansky}
\affiliation{Optoelectronics Research Centre, University of Southampton, Southampton SO17 1BJ, United Kingdom}
\author{Brian T. Kirby}
\affiliation{Tulane University, New Orleans, LA 70118 USA}
\affiliation{United States Army Research Laboratory, Adelphi, MD 20783 USA}
\author{Li Qian}
\affiliation{Dept. of Electrical and Computer Engineering, University of Toronto, 10 King's College Rd., Toronto, M5S 3G4, Canada}

\date{\today}

\begin{abstract}

Hyperentanglement, the simultaneous and independent entanglement of quantum particles in multiple degrees of freedom, is a powerful resource that can be harnessed for efficient quantum information processing. In photonic systems, the two degrees of freedom (DoF) often used to carry quantum and classical information are polarization and frequency, thanks to their robustness in transmission, both in free space and in optical fibers. Telecom-band hyperentangled photons generated in optical fibers are of particular interest because they are compatible with existing fiber-optic infrastructure, and can be distributed over fiber networks with minimal loss. Here, we experimentally demonstrate the generation of telecom-band biphotons hyperentangled in both the polarization and frequency DoFs using a periodically-poled silica fiber and observe entanglement concurrences above 0.95 for both polarization and frequency DOFs. Furthermore, by concatenating a Hong-Ou-Mandel interference test for frequency entanglement and full state tomography for polarization entanglement in a single experiment, we can demonstrate simultaneous entanglement in both the polarization and frequency DOFs.  The states produced by our hyperentanglement source can enable protocols such as dense coding and high-dimensional quantum key distribution.
\end{abstract}

\maketitle

\section{Introduction}
Exploiting the simultaneous entanglement in multiple degrees of freedom (DOFs), or hyperentanglement, of an entangled photon pair increases the dimensionality of the Hilbert space for quantum information processing. Due to its high capacity for quantum information, hyperentanglement has attracted much interest for its applications in quantum superdense coding  \cite{Barreiro2008, Hu2018,Graham2015, Chapman2020}, complete Bell-state analysis  \cite{Ciampini2016,Williams2017}, and cluster state generation  \cite{Vallone2010, Ciampini2016}. The generation of hyperentangled photons has been demonstrated in various combinations of DOFs, such as polarization and spatial modes  \cite{Barreiro2005}, polarization and time-bin  \cite{Chapman2020}, and polarization and orbital angular momentum  \cite{Zhao2019}. Here, we consider hyperentanglement in frequency and polarization DOFs. Entanglement in these two DOFs can be generated straightforwardly in fiber \cite{Chen2017} and nonlinear waveguides \cite{Martin2010} via nonlinear processes such as spontaneous parametric down-conversion (SPDC) and spontaneous four-wave-mixing (SFWM).  While these waveguide-based sources benefit from greater mode confinement (compared to their bulk crystal counterparts) and single spatial-mode emission, this latter property also limits the accessible dimensionality of the hyperentangled photon pairs generated, due to the fact that the two entangled photons are in the spatial mode \cite{Chen2020}. Beamsplitter-based techniques have been used for the probabilistic separation of the biphotons to achieve polarization-frequency (PF) hyperentanglement  \cite{Xie2015}, but the hyperentanglement is achieved only after post-selection of coincidence detection, with only 50\% probability, limiting its potential for applications such as dense coding. To generate PF hyperentanglement where entanglement is individually accessible in each DOF, we need to separate the biphotons into two spatial modes deterministically. 

\begin{figure*}[t]
\centering
\includegraphics[width=17cm]{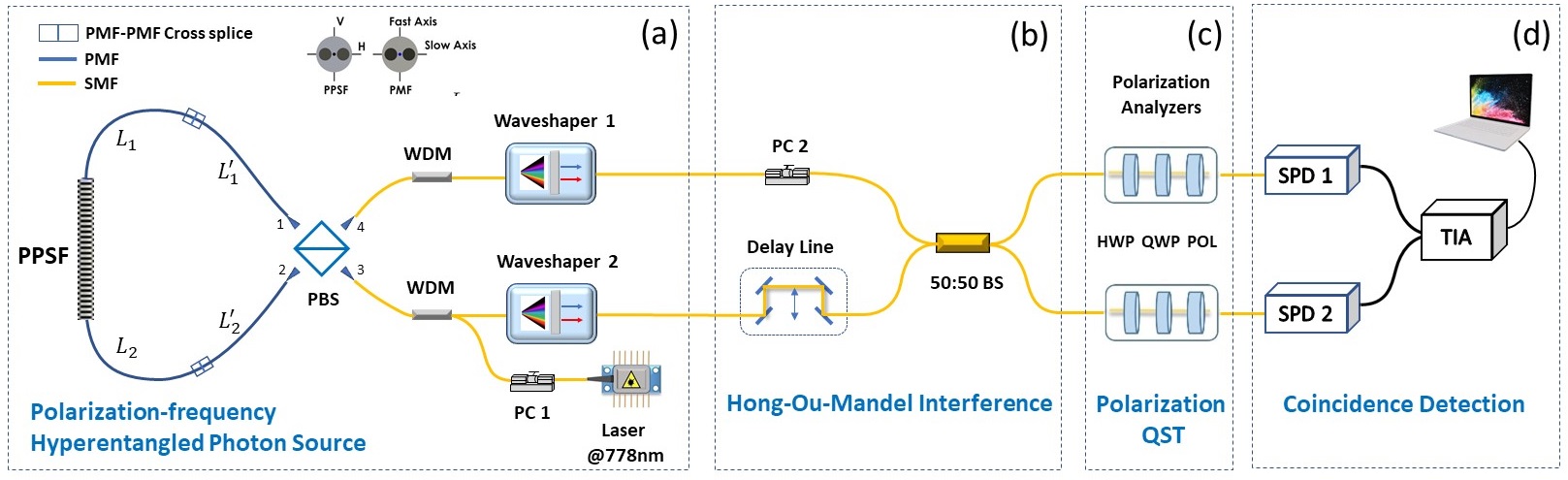}
\caption{Experimental setup of the polarization-frequency hyperentangled photon source and accompanying characterization setup. (a) The PPSF-based polarization-frequency hyperentangled photon source. PBS, polarizing beamsplitter; WDM, wavelength division demultiplexer for 780 nm/1550 nm; Waveshaper, a programmable optical filter (Finasar WaveShaper 4000X); PC, polarization controller; Blue lines are polarization-maintaining fiber (PMF), and yellow lines are single mode fiber (SMF); Laser @778nm, a wavelength tunable cw laser with its emission wavelength set to 778nm. The inset shows the correspondence of polarization modes between PPSF and PMF. (b) Experimental setup of Hong-Ou-Mandel interference. 50:50 BS, 50:50 beamspliter. (c) Polarization analyzers which are used for polarization quantum state tomography. Each polarization analyzer includes a half-wave plate (HWP), a quarter-wave plate (QWP), and a polarizer (POL). (d) Photon coincidence detection setup. SPD, single photon detector; TIA, time interval analyzer. }
\label{fig:expsetup}
\end{figure*}

In order to deterministically separate the PF hyperentangled biphotons generated from collinearly propagating biphotons without destroying the entanglement in either DOF, one can make use of biphoton interference, specifically, the anti-bunching effect on a beamsplitter or a polarizing beamsplitter  \cite{Marchildon2016, Chen2020a}.  Based on the proposed design in our recent work  \cite{Chen2020}, here we present an experimental demonstration of a PF hyperentangled photon-pair source. The photons are generated via a broadband type-II spontaneous parametric down conversion in a periodically-poled silica fiber (PPSF) \cite {Eric2013}, bidirectionally-pumped inside a Sagnac loop. Further improvements are made to the design proposed in \cite{Chen2020} which ease the requirements of birefringence compensation. We measure the polarization entanglement by performing quantum state tomography (QST) in the polarization subspace independently, as well as verify the existence of frequency entanglement by examining the spatial anti-symmetry of the biphoton states via Hong-Ou-Mandel interference (HOMI)  \cite{Fedrizzi2009}. In addition, we show that the entanglement in polarization and frequency exists simultaneously and independently in the biphotons of two spatial modes by performing polarization QST at the HOMI peak, where the photon anti-bunching effect at the HOMI peak verifies the frequency entanglement and the QST verifies the polarization entanglement. The PPSF-based PF hyperentangled photon source exhibits high entanglement qualities in both polarization and frequency DOFs. It could serve as an enabler for high-dimensional quantum information processing \cite{Wei2007, Lukens2016, Luo2019}.

\section{PPSF-based Sagnac-loop PF hyperentangled photon source}

The experimental setup of the PPSF-based PF-hyperentangled photon source is shown in Fig.\ref{fig:expsetup}(a). The PPSF is a 20-cm-long, weakly birefringent step-index silica fiber with second-order nonlinearity induced by thermal poling  \cite{Canagasabey2009}. The PPSF can be used for direct polarization entanglement generation without any compensation    \cite{Chen2017}.  The quasi-phase-matching (QPM) condition is achieved through periodic UV erasure with period $\Lambda$ = 62 $\mu$m, which results in a degeneracy wavelength of 1556 nm (192.67 THz) at 20$^o$C for biphoton generation via type-II SPDC. A pair of orthogonally polarized photons is generated by the down-conversion of a pump photon whose polarization is along the V axis of the PPSF, as is defined in the inset of  Fig.\ref{fig:expsetup}(a). The PPSF is bidirectionally pumped in the Sagnac-loop. Though the downconverted photons are broadband, for simplicity, let us consider the biphoton state of only a pair of frequency bins, which can be obtained, for example, using a suitable two-teeth frequency filter. The output state at either end of the PPSF can then be written as (see Appendix A Eq. (\ref{eq_a2})):
\begin{align}
\ket{\Psi}_X = \frac{1}{\sqrt{2}}e^{i\phi_{pX}}\left(\ket{H,\omega_s}_X\ket{V, \omega_i}_X + \ket{V, \omega_s}_X\ket{H, \omega_i}_X\right)\label{eq_singlepass}
\end{align}
where subscript $X = 1, 2$ denotes the spatial modes of the clockwise and counter-clockwise propagation directions, $\phi_{pX}$ is the constant pump phase carried over by the pump photon, $\omega_s$ and $\omega_i$ are the center angular frequencies of the signal and idler frequency bins, and we assume that $\omega_s>\omega_i$. 

The pump laser of 778 nm (Toptica DL-Pro) is first sent through a 780/1550 nm wavelength division multiplexer (WDM) before it enters into the Sagnac-loop through a polarization beamsplitter (PBS, extinction ratio $>$20 dB at both 780 and 1550 nm). The polarization of the input pump light to the PBS is controlled by a polarization controller (PC 1), such that the pump power in the clockwise and counter-clockwise directions of the Sagnac-loop is equalized. 

The PPSF is placed inside the Sagnac-loop and spliced to polarization-maintaining fibers (PMFs) of length $L_1$ and $L_2$ respectively with both their slow and fast axes aligned. Because the PMFs have large birefringence (PM1550, beat length $\sim$4mm), compensation is needed to avoid polarization entanglement decoherence due to temporal walkoff. This compensation comes in the form of two additional PMFs of length $L_1'$ and $L_2' $, which are cross-spliced to PMFs $L_1$ and $L_2$ respectively, i.e., the principal axes of $L_1'$ ($L_2'$) are rotated 90 degrees from that of $L_1$ ($L_2$). Experimentally, the length differences $|L_1-L_1'|$ and $|L_2-L_2'|$ are limited to be no more than 5 mm so that high quality entanglement in polarization and frequency DoFs will be maintained in the spectral range of interest (see Appendix C for further discussion) \cite{Chen2020, Vergyris2017}. The outputs of PMFs $L_1'$ and $L_2'$ are then aligned to the PBS, such that the pump beam is aligned to to the V-polarization of the PPSF. 

As is shown in Ref. \cite{Chen2020}, the state in Eq. (\ref{eq_singlepass}) is not truly hyperentangled because both photons in a pair are in the same spatial mode. The PBS in the Sagnac-loop is used to interfere the biphoton states and deterministically separate them at its output ports 3 and 4 to achieve hyperentanglement. Following the port 3 and 4, two 780 nm/1550 nm WDMs are used to suppress the 778 nm pump laser power. To obtain frequency bins, two programmable filters (Finisar Waveshaper 4000X) are used to generate the desired frequency bins. Assuming that the biphoton brightness is the same in ports 1 and 2 of the PBS (achieved by tuning PC 1), and $L_1 = L_1'$, $L_2 = L_2'$ (see Appendix C, for the case where these assumptions do not hold true), we may write the output biphoton state at port 3 and port 4 as (see Appendix A Eq. (\ref{eq_a5})):
\begin{align}
\ket{\Psi_{PF}} = \frac{1}{2}(\ket{H}_3\ket{V}_4) &+ e^{i\phi_{pol}}\ket{V}_3\ket{H}_4)\notag\\
\otimes&(\ket{\omega_s}_3\ket{\omega_i}_4 + \ket{\omega_i}_3\ket{\omega_s}_4)\label{eq_PF}
\end{align}
where $\phi_{pol} = \phi_{p2}-\phi_{p1}+ [(k_H(\omega_s)+k_H(\omega_i) + k_V(\omega_s) + k_V(\omega_i)](L_2- L_1)$, and $k_{H/V}(\omega)$ is the PMF's propagation constant along its slow (fast) axis at angular frequency $\omega$, and by the same convention we label it with subscript H (V). In Eq. (\ref{eq_PF}), the biphotons are in two different spatial modes and the maximally entangled states in polarization and frequency DOFs are decoupled from each other. The entanglement in each DOFs can be accessed individually and simultaneously without affecting the other, hence the generated biphoton state is PF hyperentangled.

\section{Polarization Entanglement}

\begin{figure}[t]
\centering
\includegraphics[width=8.5cm]{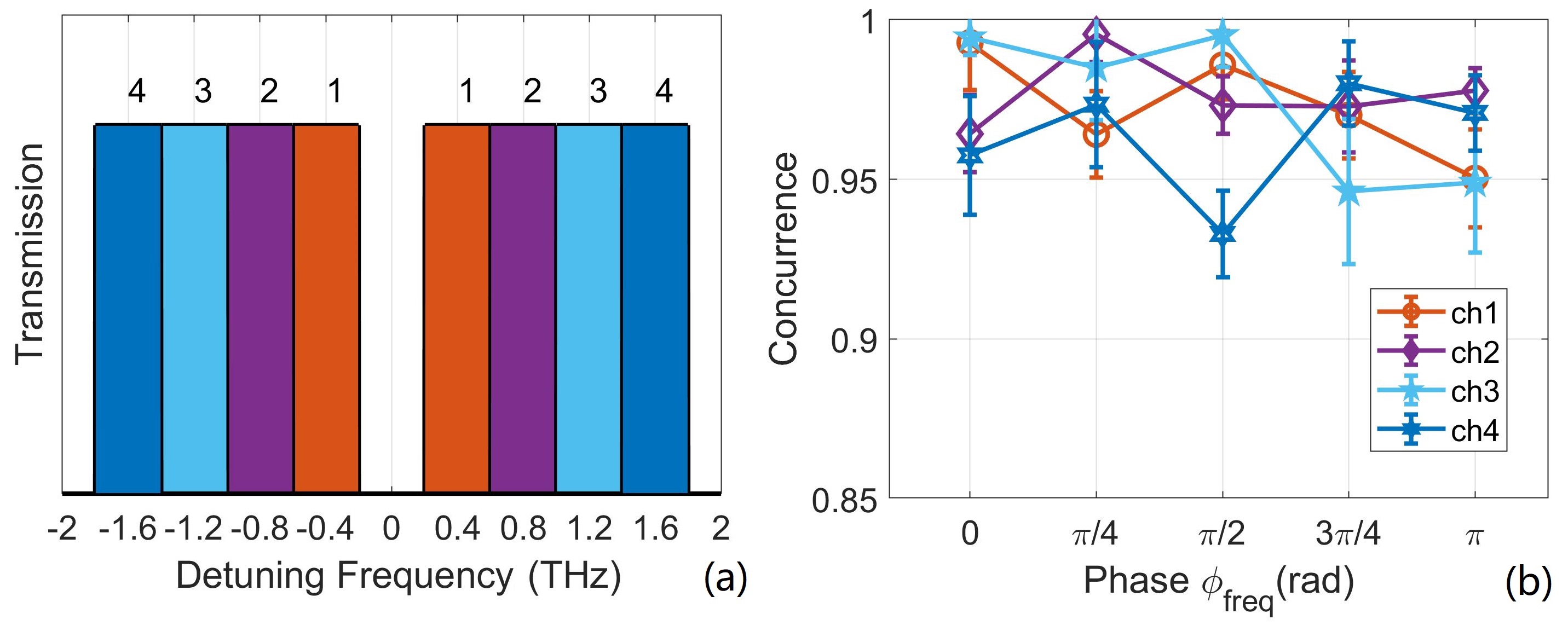}
\caption{(a) Transmission spectra of the waveshapers. (b) Concurrences measured in polarization quantum state tomography with varying spectral phase $\phi_{freq}$. }
\label{fig:waveshaperspec}
\end{figure}

We first demonstrate the polarization entanglement of the PF hyperentangled photon source. The polarization entanglement was measured with the quantum state tomography (QST) setup combining Fig.\ref{fig:expsetup}(a), (c), and (d). The hyperentangled biphoton output from Fig.\ref{fig:expsetup}(a) is directly connected to a pair of polarization analyzers (PAs, Hewlett Packard 8169A) in Fig.\ref{fig:expsetup}(c). Each PA consists of a half-wave plate, a quarter-wave plate, and a polarizer. Finally, the PAs are followed by the coincidence detection setup in Fig.\ref{fig:expsetup}(d). Several pairs of frequency bins having various detuning frequencies ($\pm0.4$ THz, $\pm0.8$ THz, $\pm1.2$ THz, and $\pm1.6$ THz, labeled as channel 1, 2, 3, and 4 respectively) from degeneracy were used for demonstration, as shown in Fig.\ref{fig:waveshaperspec}(a). Each pair of frequency bins has a two-teeth top-hat transmission spectrum with a passband width of 0.4 THz, created by the programmable waveshapers. The waveshapers can also apply a frequency-dependent phase to the photons that can be used for encoding the hyperentangled state in the frequency DoF. Note that, the frequency-dependent phase applied by the waveshapers does not affect the polarization entanglement. Before performing QST, we varied the spectral phase by setting the waveshaper 1 to apply zero phase to $\ket{\omega_i}_4$, and a phase $\phi_{freq}$ to $\ket{\omega_s}_4$. We set Waveshaper 2 to apply zero phase to both frequency bins in mode 3, such that the state from Eq. (\ref{eq_PF}) becomes (see Appendix A Eq. (\ref{eq_a5}) and Appendix B Eq. (\ref{eq_b4})):
\begin{align}
\ket{\Psi_{PF, ws}}= \frac{1}{2}(&\ket{H}_3\ket{V}_4)+ e^{i\phi_{pol}}\ket{V}_3\ket{H}_4)\notag\\
\otimes&(\ket{\omega_s}_3\ket{\omega_i}_4 + e^{i\phi_{freq}}\ket{\omega_i}_3\ket{\omega_s}_4).\label{eq_originstate}
\end{align}
We performed polarization QST with varying $\phi_{freq}$, from which we calculated the concurrence  \cite{Wootters1998} for the biphoton entanglement in the polarization DoF, as a function of $\phi_{freq}$. As is shown in Fig.\ref{fig:waveshaperspec}(b), the concurrences of the polarization entanglement are higher than 0.94 in all 4 channels and all $\phi_{freq}$ measured. As mentioned, the polarization entanglement is independent of $\phi_{freq}$ and the changes in concurrence are mainly because of the imperfect polarization alignment and the statistical error in coincidence measurements.

\section{Frequency Entanglement}
\begin{figure}[t]
\centering
\includegraphics[width=8cm]{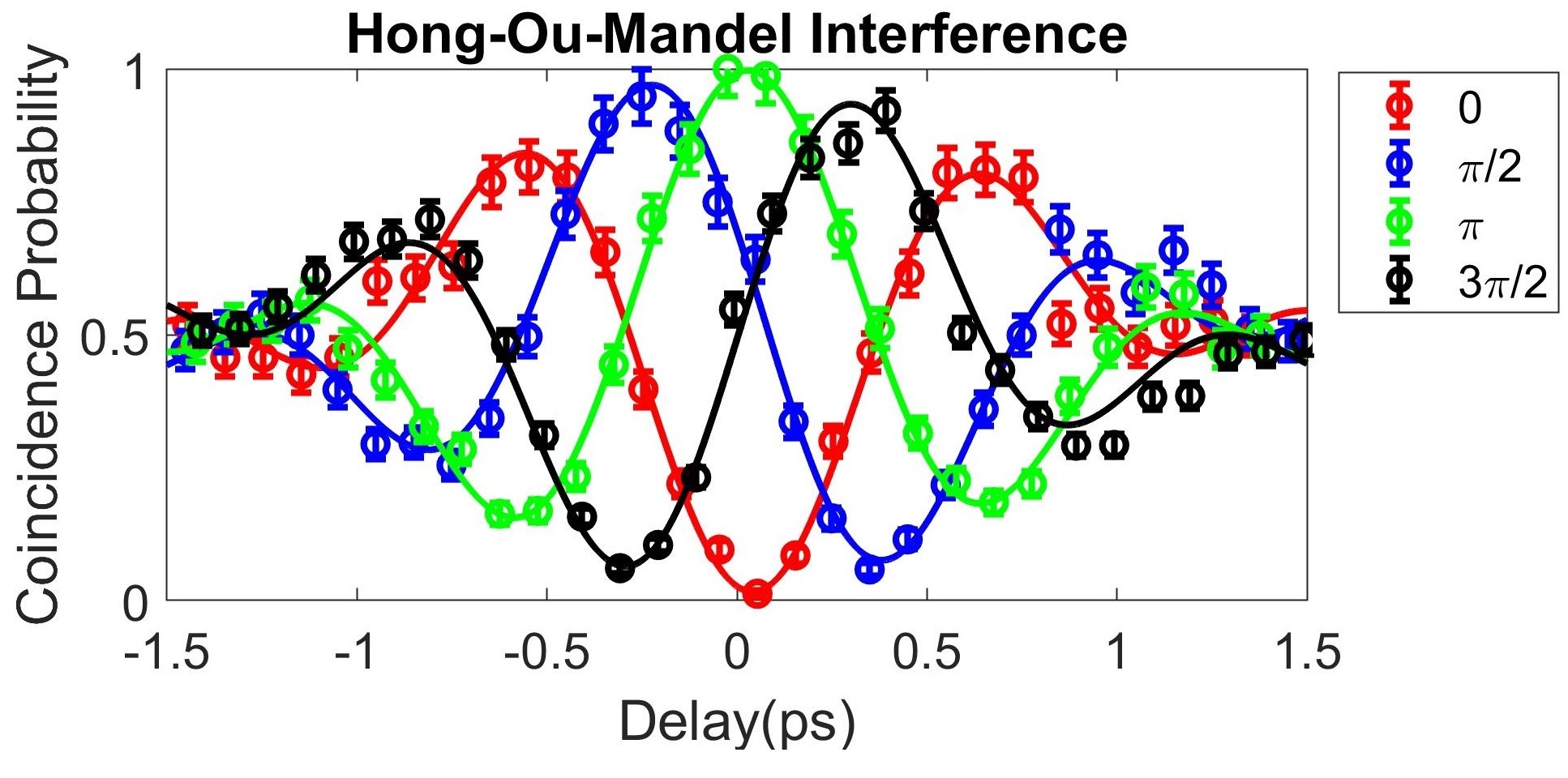}
\caption{Hong-Ou-Mandel interference of frequency-entangled photons in channel 1. The experimental interference patterns (markers) were measured under various $\phi_{freq}$ settings in the waveshapers. The simulation curves (solid lines) are the fitting curve using the model shown in Eq. (\ref{eq_b1}), where visibility $V$ of the interference and the spectral phase $\phi_{freq}$ are the coefficients of the fit. Four frequency bin entangled states with phase $\phi_{freq} = 0, \pi/2, \pi$, and $3\pi/2$ constitute two unbiased bases. }
\label{fig:homi_all}
\end{figure}

The verification of entanglement in the frequency DOF is more elaborate due to the difficulty of performing a mutually unbiased measurement in the frequency domain. While a nonlocal measurement of the coherence of frequency-entangled states is difficult without frequency conversion   \cite{Lu2018} or a time-resolved measurement   \cite{Guo2017}, the spatial beating in Hong-Ou-Mandel interference (HOMI)  \cite{Ramelow2009, Fedrizzi2009, Kaneda2019} can be used as a test of frequency entanglement. More specifically, the two-photon anti-bunching effect in HOMI is an unambiguous signature of spatially anti-symmetric entanglement in the biphoton state under test  \cite{Fedrizzi2009, Wang2006}. Considering the state given in Eq. (\ref{eq_originstate}), if the two-photon anti-bunching effect is observed in HOMI while the polarization state of the photons is spatially symmetric, then the interference peak obtained in HOMI indicates the existence of anti-symmetric entanglement in frequency DOF. 

To perform HOMI, we used an experimental setup combining Fig.\ref{fig:expsetup}(a), (b) and (d). The biphotons generated in the PF hyperentangled photon source in Fig.\ref{fig:expsetup}(a) were sent to the HOM intereferometer in Fig.\ref{fig:expsetup}(b) which consists of a polarization controller, a programmable delay line (General Photonics Inc. MDL-001), and a 50:50 beamsplitter (BS). The outputs of the BS were directly connected to the setup in Fig.\ref{fig:expsetup}(d) for coincidence detection.  Since the generated polarization-entangled biphotons arose from type-II SPDC, the polarization of each photon in a photon-pair must be orthogonal to the polarization of its twin at the output ports 3 and 4 of the PBS. To make sure that the polarization biphoton state was spatially symmetric, a polarization controller (PC 2) was used to rotate the photon in mode 4 such that the polarizations of the biphotons at the inputs of the BS were identical (in the reference frame of the BS), such that the biphoton state can be written as (see Appendix B Eq. (\ref{eq_b4})):
\begin{align}
\ket{\Psi_{PF, ws}}= \frac{1}{2}(&\ket{H}_3\ket{H}_4)+ e^{i\phi_{pol}}\ket{V}_3\ket{V}_4)\notag\\
\otimes&(\ket{\omega_s}_3\ket{\omega_i}_4 + e^{i\phi_{freq}}\ket{\omega_i}_3\ket{\omega_s}_4).\label{eq_HOMstate}
\end{align}
The biphoton state in the polarization DOF is hence spatially symmetric regardless of $\phi_{pol}$. With a spatially symmetric polarization state aligned to the BS, any observation of photon anti-bunching in HOMI will only result from the anti-symmetric entanglement in the frequency DOF. To demonstrate the effect of spatial anti-symmetry in the frequency domain, we introduce a frequency-dependent phase $\phi_{freq}$, as described in Section III. We expect to obtain a PF hyperentangled state of the form given by Eq. (\ref{eq_HOMstate}). The HOMI pattern can be further modeled with a coincidence probability $p$ as a function of the arrival time delay $\tau$ of the biphotons and the phase $\phi_{freq}$ (see Appendix B Eq. (\ref{eq_b6})):
\begin{align}
p(\tau,\phi_{freq}) = \frac{1}{2} - \frac{V}{2}\mathrm{sinc}(\delta\omega\tau)\cos(2\omega_0\tau- \phi_{freq})\label{eq_coincprob}
\end{align}
where $V$ is the visibility of the HOMI, $\delta\omega$ is the bandwidth of each frequency bin, and $\omega_0$ is the bin's center detuning frequency from the degeneracy. When $\phi_{freq} = 0$, the state shown in Eq. (\ref{eq_HOMstate}) becomes spatially symmetric and the photon-bunching effect, or a HOMI dip, is expected. In contrast, when $\phi_{freq} = \pi$, the state becomes spatially anti-symmetric and a HOMI peak is obtained. 

As is shown in Fig.\ref{fig:homi_all} where channel 1 ($\delta\omega$ = $2\pi\times$0.4 THz, $\omega_{0}$ = $2\pi\times$0.4 THz) is used for demonstration, with $\phi_{freq}$ varying from 0 to $\pi$, the HOMI changes from a HOMI dip in which biphotons are bunching, to a HOMI peak in which biphotons are anti-bunching. A visibility of greater than 96.9\% (average 98.9\%) are obtained in all interferograms by fitting Eq. (\ref{eq_coincprob}) to the experimentally measured coincidence counts. The anti-symmetry of the biphoton state which was revealed by two-photon anti-bunching in HOMI, together with the fact that the biphotons in modes 3 and 4 were identically polarized,  shows that the biphotons in modes 3 and 4 are strongly entangled in the frequency DOF  \cite{Fedrizzi2009}. In addition, Fig.\ref{fig:homi_all} shows the HOMI pattern using the biphotons with $\phi_{freq} = 0,\pi/2, \pi,$ and $3\pi/2$ which constitute a set of qubit mutually unbiased bases, illustrating the states' potential in encoding in the frequency domain.  

However, the HOMI patterns may be obtained without entanglement in the polarization domain. The polarization state that entered into the BS was not characterized concurrently with the HOMI measurement. Since the observations of polarization and frequency entanglement were performed independently, we cannot truely determine if the state is hyperentangled from these results alone. The following section will solve this problem by simultaneously characterizing both DOFs in our hyperentangled source.

\section{Simultaneous entanglement in polarization and frequency DOF}
In this section we simultaneously observe both polarization and frequency entanglement by performing polarization QST on anti-bunched photons that are the output of a HOMI experiment with a setup combining Fig.\ref{fig:expsetup}(a), (b), (c), and (d). The results of the combined measurements therefore suggest the existence of polarization-frequency hyperentanglement. 

\begin{figure}[t]
\centering
\includegraphics[width=8.5cm]{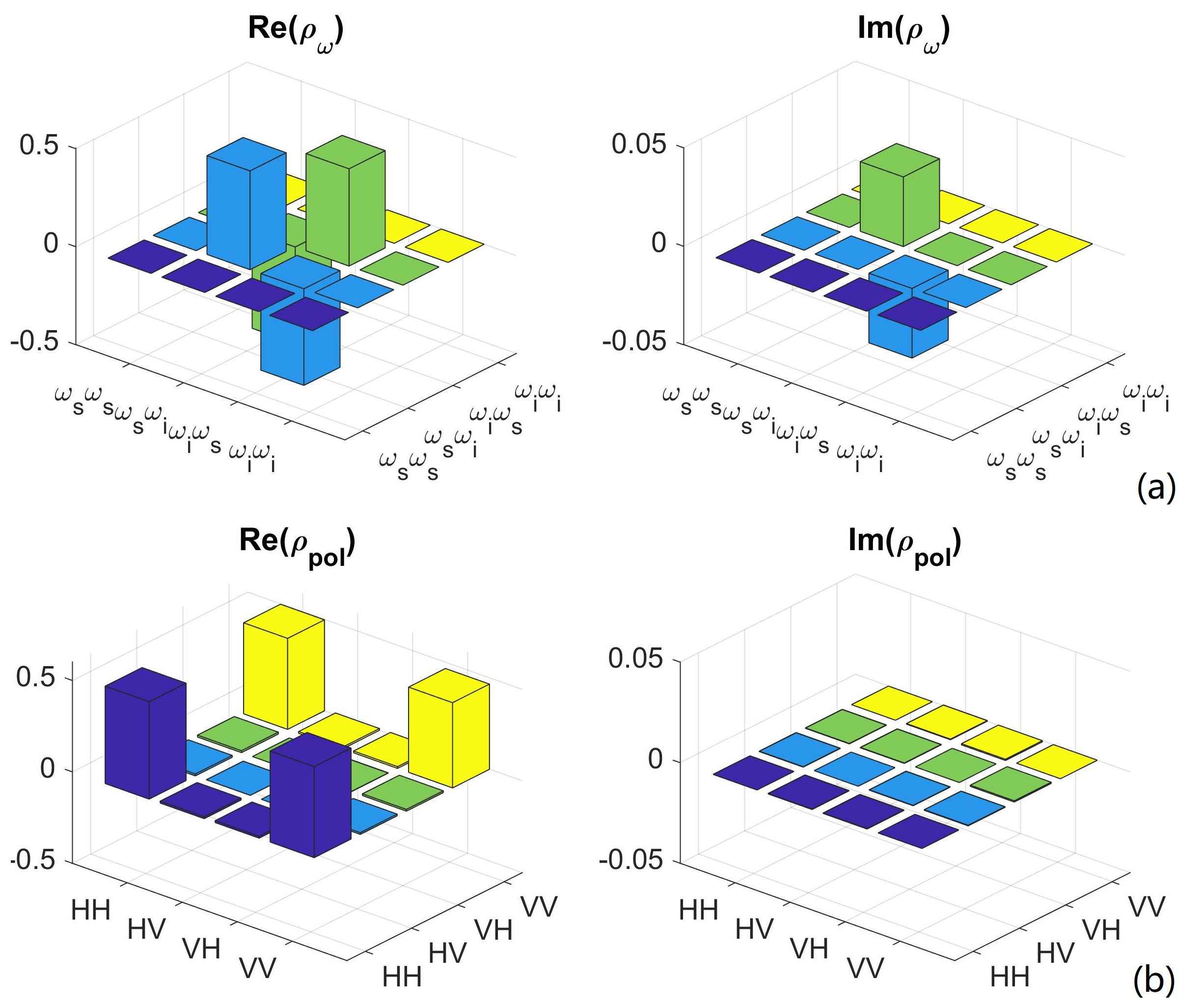}
\caption{The real and imaginary parts of the reduced density matrices of the PF hyperentangled state which are reconstructed from the experimental data in: (a) Frequency domain $\rho_\omega$; (b) Polarization domain $\rho_{pol}$}
\label{fig:dms}
\end{figure}

We demonstrate the simultaneous entanglement in polarization and frequency DOFs using channel 1. With $\phi_{freq}$ set to $\pi$, a visibility V = $98.8\pm1.7\%$ and a experimental measured $\phi_{freq} = 3.07\pm0.10$ rad are obtained by fitting Eq. (\ref{eq_coincprob}) to the experimentally measured coincidence counts. We may estimate the reduced density matrix within the frequency bins subspace  \cite{Ramelow2009, Kaneda2019} by writing the density matrix in the computational basis, $\{\ket{\omega_s}_3\ket{\omega_s}_4, \ket{\omega_s}_3\ket{\omega_i}_4, \ket{\omega_i}_3\ket{\omega_s}_4, \ket{\omega_i}_3\ket{\omega_i}_4\}$:
\begin{align}
\rho_\omega = \left(\begin{matrix}
0 & 0 & 0 & 0\\
0 & p_\omega & \frac{V}{2}e^{i\phi_{freq}} & 0\\
0 & \frac{V}{2}e^{-i\phi_{freq}} & 1-p_\omega & 0\\
0 & 0 & 0 & 0
\end{matrix}\right)\notag
\end{align}
where $p_\omega$ is the probability of $\ket{\omega_s}_3\ket{\omega_i}_4$, and the density matrix elements involving $\ket{\omega_s}_3\ket{\omega_s}_4$ and $\ket{\omega_i}_3\ket{\omega_i}_4$ are all assumed to be zero due to the constraints imposed by energy conservation in a SPDC process pumped by a narrow-band cw laser\cite{Ramelow2009, Chen2020a}. We obtained $p_\omega = 0. 504 \pm 0.022$ by measuring the ratio of coincidence count rates of $\ket{\omega_s}_3\ket{\omega_i}_4$ and $\ket{\omega_i}_3\ket{\omega_s}_4$ directly. The reconstructed density matrix in the frequency domain is shown in Fig.\ref{fig:dms}(a), and a concurrence of $C(\rho_\omega) \approx 0.988$ in the frequency subspace is obtained. We then fix the delay to $\tau = 0$ and perform polarization QST on the anti-bunching biphotons at the output of the BS. The measured reduced density matrix $\rho_{pol}$ in polarization domain [Fig.\ref{fig:dms}(b)] yields a polarization concurrence of $C(\rho_{pol}) = 0.986\pm0.019$. 

We may further infer the entanglement properties of the global state in both polarization and frequency DOF via semidefinite programming (SDP) in the same fashion as Ref. \cite{Chen2020a}, where only a weaker assumption of $\bra{\omega_s\omega_s}\rho_\omega\ket{\omega_s\omega_s}=\bra{\omega_i\omega_i}\rho_\omega\ket{\omega_i\omega_i}=0$ needs to be made. The fidelity of the reduced polarization state to a Bell state $\ket{\Phi_p^+}=\frac{1}{\sqrt{2}}(\ket{H}\ket{H}+\ket{V}\ket{V})$ is calculated to be $F_p = 0.997$, and the lower bound of fidelity of the reduced frequency state to $\ket{\Psi_\omega^-} = \frac{1}{\sqrt{2}}(\ket{\omega_s}\ket{\omega_i}-\ket{\omega_i}\ket{\omega_s})$ is provided by the visibility of HOMI $F_\omega\geq0.988$. Given the fidelities obtained in each subspace, the lower bound of the fidelity $F_{p\omega}$ of the global state to $\ket{\Phi_p^+}\otimes\ket{\Psi_\omega^-}$ is calculated to be $F_{p\omega}\geq0.985$ by SDP. This further certifies the quality and the hyperentanglement in our PPSF-based source. Therefore, we have demonstrated successive and simultaneous entanglement characterization in both the polarization and frequency DOF of our hyperentanglement source and these results are consistent with the existence of high quality PF hyperentanglement.

\section{Conclusions}
In this work, we have demonstrated a PF hyperentangled photon-pair source based on the PPSF technology. The PF hyperentangled photon-pairs are deterministically separated into two spatial modes, which provides more accessible dimensionalities than the collinear non-degenerate entangled photon source  \cite{Chen2020}. We experimentally verified the simultaneous entanglement in the polarization and frequency DOFs through polarization QST and the two-photon anti-bunching effect in HOMI. Our method, using type-II SPDC in only one PPSF, removes the requirement for two identical nonlinear media, as is the case in Ref.\cite{Chen2020a, Marchildon2016}. The Sagnac loop in our source can potentially be implemented with all fibers and fiber-pigtailed micro-optic components, without free-space coupling or alignment. With further development, we anticipate that the PF hyperentangled state can be used in quantum applications such as cluster state generation, high-dimensional quantum communication  \cite{Ciampini2016}, and photonic quantum computation  \cite{Lukens2016}.

\section*{Appendix A: PF Hyperentangled State Generation in PPSF-based Sagnac-Loop Biphoton Source}
In the following we will show the calculation of the output state of the Sagnac-loop hyperentangled photon source. We may write the biphoton state of a pair of frequency bins generated via type-II SPDC at the output ends of the PPSF inside the Sagnac-loop  \cite{Chen2017}:
\begin{align}
\ket{\Psi^+}_X = \frac{1}{\sqrt{2}}e^{i\phi_{pX}}\iint f(\omega_s, \omega_i)\big(\ket{H,\omega_s}_X\ket{V, \omega_i}_X \notag\\
+ \ket{V, \omega_s}_X\ket{H, \omega_i}_X\big)\mathrm{d}\omega_s\mathrm{d}\omega_i\tag{A1}\label{eq_AppendixAsinglepass}
\end{align}
where $\ket{P, \omega}_X = \ket{P}_X\otimes\ket{\omega}_X$ denotes a single photon state at frequency $\omega$, in $P = H$ or $V$ polarization, and in  $X$ = 1, or 2 spatial mode (1 for clockwise and 2 for counter-clockwise propagation direction), $\phi_{pX}$ is the constant pump phase carried over by the pump photon, $\omega_s$ and $\omega_i$ are the center angular  frequencies of the signal and idler frequency bins, and we assume that $\omega_s>\omega_i$. $f(\omega_s, \omega_i)$ is the joint spectral amplitude function of the biphotons, where we have implicitly assumed that it is identical for both $\ket{H,\omega_s}_X\ket{V, \omega_i}_X$ and $\ket{V,\omega_s}_X\ket{H, \omega_i}_X$ because of the low group birefringence in PPSF  \cite{Chen2017}. For simplicity, we may consider frequency bins with narrow linewidth, and write the state in the form of discretized frequency bins:
\begin{align}
\ket{\Psi}_X \rightarrow \frac{1}{\sqrt{2}}e^{i\phi_{pX}}\big(\ket{H,\omega_s}_X\ket{V, \omega_i}_X + \ket{V, \omega_s}_X\ket{H, \omega_i}_X\big)\tag{A2}\label{eq_a2}
\end{align}
The state above is also shown as Eq. (\ref{eq_singlepass}) in the main text. 

The biphotons state generated by the PPSF will propagate through PMF $L_X$, a PMF to PMF cross-splice, and PMF $L_X'$ as shown in Fig.\ref{fig:expsetup}(a). Assuming that the propagation constant of PMF as a function of angular frequency $\omega$ at its slow (fast) axis is $k_{H/V}(\omega)$, the biphoton state arrives at the PBS will become:
\begin{widetext}
\begin{align}
\ket{\Psi}_X\rightarrow\frac{1}{\sqrt{2}}e^{i\phi_{pX}}\Big(\ket{H,\omega_s}_X\ket{V, \omega_i}_Xe^{i\{[k_H(\omega_s)+k_V(\omega_i)]L_X+[k_V(\omega_s)+k_H(\omega_i)]L_X'\}}\notag\\
 + \ket{V, \omega_s}_X\ket{H, \omega_i}_Xe^{i\{[k_V(\omega_s)+k_H(\omega_i)]L_X+[k_H(\omega_s)+k_V(\omega_i)]L_X'\}}\Big)\tag{A3}
\end{align}
\end{widetext}
At the PBS, all of the H-polarized photons are transmitted, and all of the V-polarized photons are reflected, i.e. $\ket{H}_1\rightarrow\ket{H}_3$, $\ket{H}_2\rightarrow\ket{H}_4$, $\ket{V}_1\rightarrow\ket{V}_4$, $\ket{V}_2\rightarrow\ket{V}_3$. Note that the biphotons that arrive at the PBS shall have the same brightness for $X$ = 1 and 2, which can be experimentally achieved by properly aligning the polarization of the pump laser using polarization controller 1 in Fig.\ref{fig:expsetup}(a). The cw-laser pump is assumed to have a long coherence length such that the biphotons will be coherently superposed at the PBS. The output state at modes 3 and 4 of the PBS will be:
\begin{widetext}
\begin{align}
\frac{1}{\sqrt{2}}(\ket{\Psi}_1 + \ket{\Psi}_2)\xrightarrow{PBS}\frac{1}{2}\bigg[&e^{i\phi_{p1}}\ket{H,\omega_s}_3\ket{V, \omega_i}_4e^{i\{[k_H(\omega_s)+k_V(\omega_i)]L_1+[k_V(\omega_s)+k_H(\omega_i)]L_1'\}}\notag\\
&e^{i\phi_{p1}}\ket{V,\omega_s}_4\ket{H, \omega_i}_3e^{i\{[k_V(\omega_s)+k_H(\omega_i)]L_1+[k_H(\omega_s)+k_V(\omega_i)]L_1'\}}\notag\\
&e^{i\phi_{p2}}\ket{H,\omega_s}_4\ket{V, \omega_i}_3e^{i\{[k_H(\omega_s)+k_V(\omega_i)]L_2+[k_V(\omega_s)+k_H(\omega_i)]L_2'\}}\notag\\
&e^{i\phi_{p2}}\ket{V,\omega_s}_3\ket{H, \omega_i}_4e^{i\{[k_V(\omega_s)+k_H(\omega_i)]L_2+[k_V(\omega_s)+k_H(\omega_i)]L_2'\}}\bigg]\tag{A4}\label{eq_A4}
\end{align} 
\end{widetext}
Experimentally, we carefully choose the length of PMFs such that $L_1\approx L_1'$ and $L_2\approx L_2'$. Assuming that the source setup is ideal, $L_1= L_1'$, and $L_2= L_2'$, we can easily factor out the common phases. Reorganizing the Eq. (\ref{eq_A4}), we obtain:
\begin{align}
\ket{\Psi_{PF}} = \frac{1}{2}(\ket{H}_3\ket{V}_4 &+ e^{i\phi_{pol}}\ket{V}_3\ket{H}_4)\notag\\
\otimes&(\ket{\omega_s}_3\ket{\omega_i}_4 + \ket{\omega_i}_3\ket{\omega_s}_4)\label{eq_a5}\tag{A5}
\end{align}
where $\phi_{pol} = \phi_{p2}-\phi_{p1}+ [(k_H(\omega_s)+k_H(\omega_i) + k_V(\omega_s) + k_V(\omega_i)](L_2- L_1)$. $\phi_{pol}$ depends only on the length and the dispersion property of the fiber. If the length of the fiber is not varying and $L_1 \approx L_2$, $\phi_{pol}$ is approximately a constant phase. Eq. (\ref{eq_a5}) is obviously a polarization-frequency hyperentangled state and it is also shown in the main text as Eq. (\ref{eq_PF})

\section*{Appendix B: Hong-Ou-Mandel Interference in PPSF-based hyperentangled Photon Source}
In the Hong-Ou-Mandel interference (HOMI) experiment described in Section IV, a polarization controller (PC 2 in Fig.\ref{fig:expsetup}(b)) is used to rotate the polarization of the photons in mode 4 to their orthogonal polarization. Therefore, following Eq. (\ref{eq_a5}), the biphoton state that arrive at the 50:50 beamsplitter in Fig.\ref{fig:expsetup}(b) becomes:
\begin{align}
\ket{\Psi_{PF, rot}} = \frac{1}{2}(\ket{H}_3\ket{H}_4 &+ e^{i\phi_{pol}}\ket{V}_3\ket{V}_4)\notag\\
\otimes&(\ket{\omega_s}_3\ket{\omega_i}_4 + \ket{\omega_i}_3\ket{\omega_s}_4)\label{eq_b1}\tag{B1}
\end{align}
The beamsplitter in HOMI is polarization-independent, and the biphoton interference on it will only occur when the input photons are in the same polarization mode. Therefore, the HOMI with PF hyperentangled photon state in Eq. (\ref{eq_b1}) can be understood as the superposition of two HOMIs with the biphotons associated with $\ket{H}_3\ket{H}_4$ and $\ket{V}_3\ket{V}_4$ respectively. The overall HOMI pattern will be insensitive to $\phi_{pol}$. In fact, even if the biphoton polarization state is in a mixed state, we can still observe HOMI as long as the two photons in a pair are identically polarized.

To derive the HOMI pattern, we need to write the biphoton state in terms of the spectral amplitude function, which is similar to what is shown in Eq. (\ref{eq_AppendixAsinglepass}):
\begin{align}
\ket{\Psi_{PF}}_{HOM}&= \frac{1}{2}(\ket{H}_3\ket{H}_4 + e^{i\phi_{pol}}\ket{V}_3\ket{V}_4)\otimes\notag\\
\iint_{\omega_s>\omega_i}\mathrm{d}\omega_s\mathrm{d}\omega_if(&\omega_s, \omega_i)(\ket{\omega_s}_3\ket{\omega_i}_4 + \ket{\omega_i}_3\ket{\omega_s}_4)\label{eq_b2}\tag{B2}
\end{align}
We will use a top-hat two frequency bins filter programmed by the waveshapers (as shown in Fig.\ref{fig:waveshaperspec}(a)). We assume that each frequency bin has a bandwidth of $\delta\omega$, and its center detuning angular frequency from the degeneracy frequency $\omega_p/2$ is $\pm\omega_0$ ($\omega_0>0$), where a cw-laser pump of angular frequency $\omega_p$ is used. Because of energy conservation in SPDC and the assumption of a cw narrowband pump laser, we have $\omega_p = \omega_s + \omega_i$ and rewrite the state:
\begin{align}
\ket{\Psi_{PF}}_{HOM}&\propto\iint_{\omega_s>\omega_i} (\ket{\omega_s}_3\ket{\omega_i}_4 + \ket{\omega_i}_3\ket{\omega_s}_4)\notag\\
&\quad f(\omega_s, \omega_i)\delta(\omega_s+\omega_i -\omega_p)\mathrm{d}\omega_s\mathrm{d}\omega_i\notag\\
&\propto\int f(\omega)\ket{\omega_p/2 + \omega}_3\ket{\omega_p/2-\omega}_4\mathrm{d}\omega\notag
\end{align}
where $\omega$ is a dummy variable and $f(\omega)$ is the filter transmission function. We omit the polarization state here because $\ket{H}_3\ket{H}_4$ and $\ket{V}_3\ket{V}_4$ associate to identical frequency state and they will not interfere with each other. The filter transmission function $f(\omega)$ can be written as:
\begin{equation}
f(\omega) = \left\{\begin{aligned}
&\frac{1}{\sqrt{2\delta\omega}}, \quad \omega\in[\omega_0 - \frac{\delta\omega}{2},\omega_0 + \frac{\delta\omega}{2}]\\ 
&\quad\quad\quad\quad\quad\quad\cup [-\omega_0 - \frac{\delta\omega}{2},-\omega_0 + \frac{\delta\omega}{2}] \\
&0, \quad\quad\quad \mathrm{otherwise} \\
\end{aligned}\right.\notag
\end{equation}
which satisfies $\int |f(\omega)|^2\mathrm{d}\omega = 1$. The HOMI pattern, or the coincidence probability $p$ as a function of delay $\tau$ is given by  \cite{Branczyk2017}:
\begin{align}
p(\tau) &= \frac{1}{2} - \frac{1}{2}\int\mathrm{d}\omega f^*(-\omega)f(\omega)e^{2\omega\tau}\notag\\
& = \frac{1}{2} - \frac{1}{2}\mathrm{sinc}(\delta\omega\tau)\cos(2\omega_0\tau)\label{eq_b3}\tag{B3}
\end{align}
where the function $\mathrm{sinc}(x) = \sin(x)/x$.

The waveshapers can also be used to apply a frequency-dependent phase to the biphotons. Without loss of generality, we consider the following waveshaper setup: (a), the transmission spectra of the two waveshapers are $|f(\omega)|^2$; (b), the waveshaper 1 is set to apply no phase to $\ket{\omega_i}_4$, and set to apply phase $\phi_{freq}$ to $\ket{\omega_s}_4$, while the waveshaper 2 will apply no phase to any photons in mode 3. The state in Eq. (\ref{eq_b2}) that arrives at the beamsplitter becomes:
\begin{align}
\ket{\Psi_{PF}}_{HOM}&= \frac{1}{2}(\ket{H}_3\ket{H}_4 + e^{i\phi_{pol}}\ket{V}_3\ket{V}_4)\otimes\notag\\
\iint_{\omega_s>\omega_i}f(&\omega_s, \omega_i)\mathrm{d}\omega_s\mathrm{d}\omega_i(\ket{\omega_s}_3\ket{\omega_i}_4 + e^{i\phi_{freq}}\ket{\omega_i}_3\ket{\omega_s}_4).\label{eq_b4}\tag{B4}
\end{align} 
The state in Eq. (\ref{eq_b4}) rewritten in the form of discretized frequency bins is also shown as Eq. (\ref{eq_HOMstate}) in the main text. Following the same calculation procedures for Eq. (\ref{eq_b3}), we rewrite the transmission spectrum function:
\begin{equation}
f(\omega, \phi_{freq}) = \left\{\begin{aligned}
&\frac{1}{\sqrt{2\delta\omega}}, \quad\quad\quad \omega\in[\omega_0 - \frac{\delta\omega}{2},\omega_0 + \frac{\delta\omega}{2}],  \\
& \frac{1}{\sqrt{2\delta\omega}}e^{i\phi_{freq}}\quad \omega\in[-\omega_0 - \frac{\delta\omega}{2},-\omega_0 + \frac{\delta\omega}{2}],\\
&0, \quad\quad\quad\quad\quad \mathrm{otherwise}, \\
\end{aligned}\right.\notag
\end{equation}
resulting in a coincidence probability as a function of delay $\tau$ and phase $\phi_{freq}$:
\begin{align}
p(\tau,\phi_{freq}) = \frac{1}{2} - \frac{1}{2}\mathrm{sinc}(\delta\omega\tau)\cos(2\omega_0\tau- \phi_{freq})\label{eq_b5}\tag{B5}
\end{align}
In practice the interference might not be perfect and a visibility $V$ is used to quantify the interference quality. We may fit the experimental data with the following expression:
\begin{align}
p_{exp}(\tau,\phi_{freq}) = \frac{1}{2} - \frac{V}{2}\mathrm{sinc}(\delta\omega\tau)\cos(2\omega_0\tau- \phi_{freq})\label{eq_b6}\tag{B6}
\end{align}  
The coincidence probability $p(\tau,\phi_{freq})$ consists of an envelope $\mathrm{sinc}$ function modulated by a cosine function. Two particular cases are of most interest: (a), $p(\tau = 0,\phi_{freq} = 0) = 0$ indicates that no coincidence is detected, which is the case of photon-bunching effect. A HOMI dip will be observed. (b), $p(\tau = 0, \phi_{freq} = \pi) = 1$ indicates that the coincidence probability is maximized, which is the case of two-photon anti-bunching effect. The observance of an HOMI peak indicates the existence of anti-symmetric entanglement  \cite{Fedrizzi2009}.

\section*{Appendix C: Entanglement Quality Degradation due to Non-ideal Experimental Parameters}
In Appendix A, we assumed that the pump power mismatch, splice angle error, and PMF length mismatch were negligible in order to simplify the biphoton output state to that given in Eq. (\ref{eq_a5}). In this Appendix, we consider the case where these three parameters are non-negligible and use the full Eq. (\ref{eq_A4}) to analyze the impact of each parameter on the biphoton polarization entanglement.

The first parameter, the pump power mismatch, $p$, arises when the pump light entering port 3 of the PBS is not evenly split into ports 1 and 2. In order to evenly split the pump power, the pump beam must be linearly polarized at $45^\circ$ relative to the $H$ and $V$ polarizations. This mismatch parameter causes the concurrence to degrade by a constant factor across all measurements. We found that the magnitude of this degradation was relatively negligible, with $<2\%$ degradation for a $10\%$ mismatch.

\begin{figure}[htbp]
    \centering 
    \includegraphics[width=0.75\linewidth]{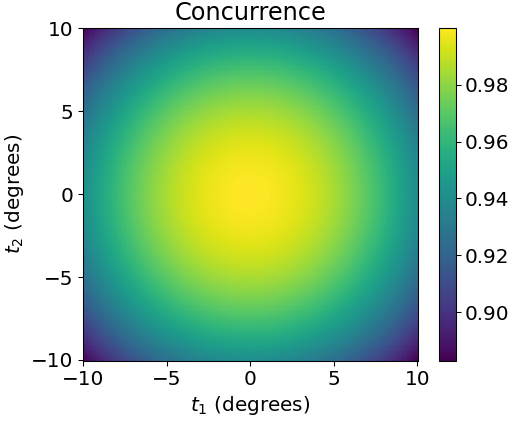}
    \caption{Simulation of the maximum degradation of polarization entanglement as a function of the splice angle error for the setup given in Fig. \ref{fig:expsetup} combining sections (a), (b), and (d), and Waveshaper filter bandwidth set to 1nm. $t_1$ and $t_2$ are the splice errors for $L_1$ and $L_2$ respectively, and the total error $t$ can be found by adding the two errors in quadrature.}
    \label{fig:conc_narrowbandfilter_angle_mismatch}
\end{figure}

The second parameter, the splice angle error, $t$, arises when the major axes of the PMF fibers are not precisely aligned at the cross splices or when the PMF is not precisely aligned to the PPSF. Each of these splices have an individual error, and through simulation we found that the total magnitude all of these individual errors can be represented by the total splice angle error $t$, as shown in Fig. \ref{fig:conc_narrowbandfilter_angle_mismatch}. The degradation due to this parameter is also relatively small, with $<3\%$ degradation for a $5^\circ$ error.

\begin{figure}[t]
    \centering 
    \includegraphics[width=0.75\linewidth]{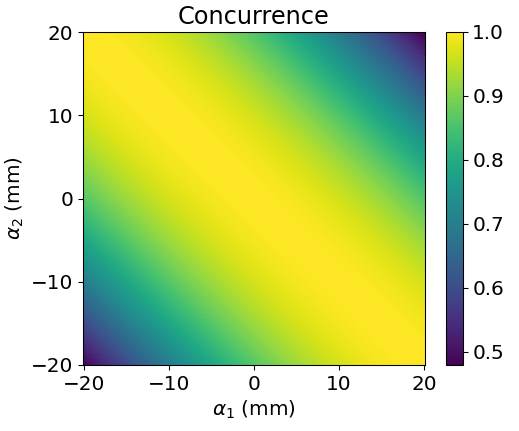}
    \caption{Simulation of the degradation of polarization entanglement at 3.3THz detuning as a function of the PMF length mismatch for the setup given in Fig. \ref{fig:expsetup} combining sections (a), (b), and (d), and Waveshaper filter bandwidth set to 1nm. The length mismatch parameters, $\alpha_1$ and $\alpha_2$, are defined such that $L_1 + \alpha_1 = L_1'$ and $L_2 +\alpha_2 = L_2'$. Note that a finite mismatch in one leg of the Sagnac loop, $\alpha_1$, can be compensated by an opposite mismatch in the other leg, $\alpha_2$.}
    \label{fig:conc_narrowbandfilter_length_mismatch}
\end{figure}

The third parameter, the PMF length mismatch, $\alpha$, arises when the lengths of the PMFs in the Sagnac loop are not identical. Letting $ L_1 + \alpha_1 = L_1'$, and $ L_2 +\alpha_2 = L_2'$, the magnitude of this degradation increases rapidly when the mismatch increases, with $2\%$ degradation for a $2$mm mismatch, $10\%$ degradation for a $5$mm mismatch, and over $30\%$ degradation for a $10$mm mismatch, as shown in Fig. \ref{fig:conc_narrowbandfilter_length_mismatch}. This is because the high birefringence of the PMFs cause significant decoherence if the length mismatch is non-ideal. Note that it is possible for the mismatch from one leg of the Sagnac loop, $\alpha_1$, to compensate for the mismatch on the other leg, $\alpha_2$. A secondary length mismatch parameter $ L_1 + \beta = L_2$ was also considered, however it was found to have a negligible effect on the concurrence.

\begin{figure}[htbp]
    \centering 
    \includegraphics[width=0.75\linewidth]{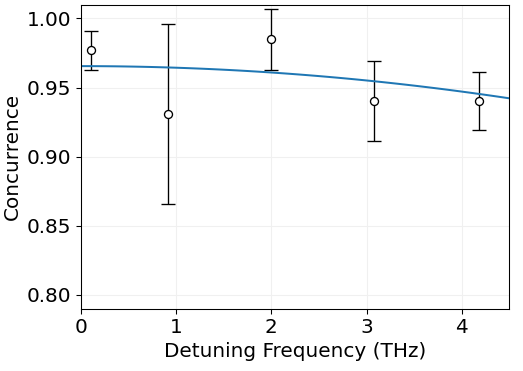}
    \caption{Fit of simulated concurrence with degradation to experimental QST data using the setup shown in Fig. \ref{fig:expsetup} combining sections (a), (b), and (d), and Waveshaper filter bandwidth set to 1nm.}
    \label{fig:conc_narrowbandfilter_expfit}
\end{figure}

\begin{figure}[htbp]
    \centering 
    \includegraphics[width=0.75\linewidth]{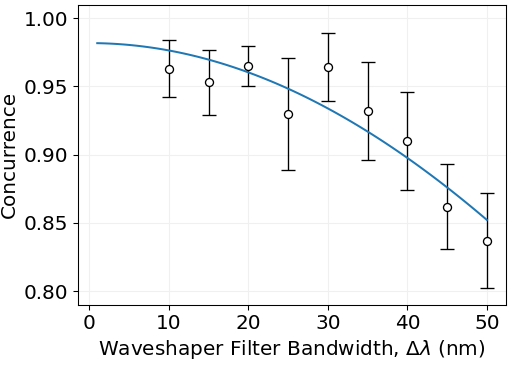}
    \caption{Fit of simulated concurrence with degradation to experimental QST data using the setup shown in Fig. \ref{fig:expsetup} combining sections (a), (b), and (d), and variedw Waveshaper filter bandwidth centered at the degeneracy point.}
    \label{fig:conc_finitebw_filter_expfit}
\end{figure}

Finally, we performed a series of QST experiments and fit the above 3 simulations to approximate the actual values of $p$, $t$, and $\alpha$. The experimental setup is the same as that shown Fig. \ref{fig:expsetup} with sections (a), (c), and (d). A reasonable fit is achieved for the values $p \leq 8\%$, $t \leq 6^\circ$, and $\alpha \leq 3$mm, as shown in Fig. \ref{fig:conc_narrowbandfilter_expfit} and Fig. \ref{fig:conc_finitebw_filter_expfit}. If, on the other hand, we simulate each parameter independently while letting the other two be 0, then we arrive at the bounds $p \leq 15\%$, $t \leq 8^\circ$, and $\alpha \leq 5$mm.


\newpage 
\thispagestyle{empty}
\mbox{}
\newpage
\bibliography{ThesisRef}

\end{document}